\documentclass[twocolumn,showpacs,preprintnumbers,amsmath,amssymb, superscriptaddress, prb]{revtex4}
\usepackage{graphicx}
\usepackage{dcolumn}
\usepackage{bm}
\usepackage{epstopdf}

\usepackage[hyperfigures=true,bookmarksnumbered=true,bookmarksopen=true,bookmarks=true,linkcolor=blue,urlcolor=blue,citecolor=blue,colorlinks=true,pdfhighlight=/O,pdfstartview={XYZ null null 1}]{hyperref}

\begin{document}

\title{Lattice effects in the quasi-two-dimensional valence-bond-solid Mott insulator EtMe$_3$P[Pd(dmit)$_2$]$_2$}

\author{Rudra Sekhar Manna}
\email[]{rudraiitb@gmail.com}
\affiliation{Physics Institute, Goethe University Frankfurt(M), D-60438 Frankfurt(M), Germany}

\author{Mariano de Souza}
\affiliation{Physics Institute, Goethe University Frankfurt(M), D-60438 Frankfurt(M), Germany}\affiliation{Instituto de Geoci\^encias e Ci\^encias Exatas - IGCE, Unesp - Univ  Estadual Paulista, Departamento de F\'isica, Cx.\,Postal 178, 13506-970 Rio Claro (SP), Brazil}

\author{Reizo Kato}
\affiliation{Condensed Molecular Materials Laboratory, RIKEN, 2-1 Hirosawa, Wako-shi, Saitama, 351-0198, Japan}

\author{Michael Lang}
\affiliation{Physics Institute, Goethe University Frankfurt(M), D-60438 Frankfurt(M), Germany}

\date{\today}


\begin{abstract}

The organic charge-transfer salt EtMe$_3$P[Pd(dmit)$_2$]$_2$ is a quasi-two-dimensional Mott insulator with localized spins $S$ = 1/2 residing on a distorted triangular lattice. Here we report measurements of the uniaxial thermal expansion coefficients $\alpha_i$ along the in-plane $i$ = $a$- and $c$-axis as well as along the out-of-plane $b$-axis for temperatures 1.4\,K $\leq T \leq$ 200\,K. Particular attention is paid to the lattice effects around the phase transition at $T_{VBS}$ = 25\,K into a low-temperature valence-bond-solid phase and the paramagnetic regime above where effects of short-range antiferromagnetic correlations can be expected. The salient results of our study include (i) the observation of strongly anisotropic lattice distortions accompanying the formation of the valence-bond-solid, and (ii) a distinct anomaly in the thermal expansion coefficients in the paramagnetic regime around 40\,K. Our results demonstrate that upon cooling through $T_{VBS}$ the in-plane $c$-axis, along which the valence bonds form, contracts while the second in-plane $a$-axis elongates by the same relative amount. Surprisingly, the dominant effect is observed for the out-of-plane $b$-axis which shrinks significantly upon cooling through $T_{VBS}$. The pronounced anomaly in $\alpha_i$ around 40\,K is attributed to short-range magnetic correlations. It is argued that the position of this maximum, relative to that in the magnetic susceptibility around 70\,K, speaks in favor of a more anisotropic triangular-lattice scenario for this compound than previously thought.

\end{abstract}

\maketitle

\section{Introduction}
\begin{figure}[!htb]
\centering
\includegraphics[width=0.75\columnwidth]{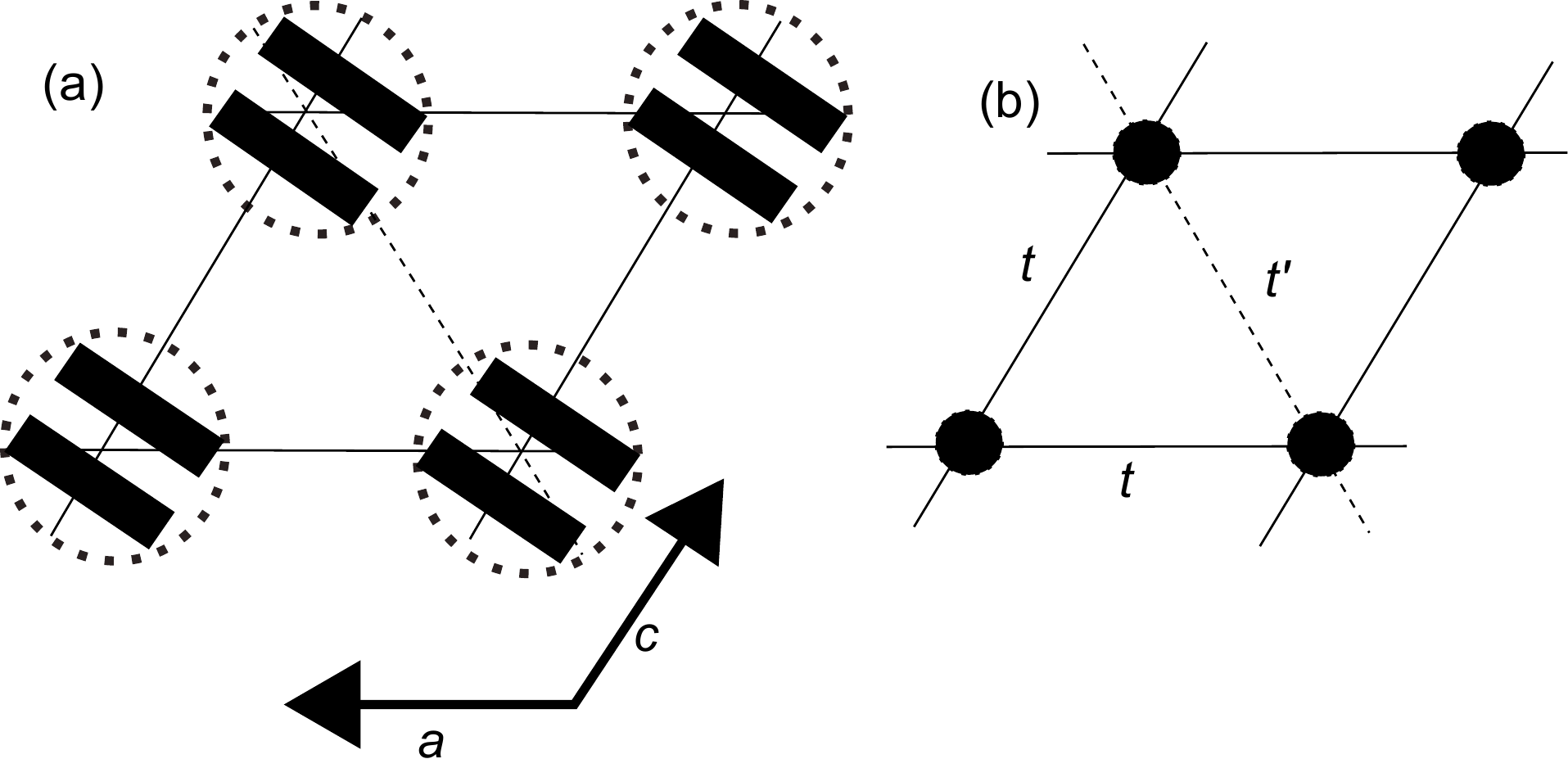}
\caption{\label{structure} (color online). (a) Schematic representation of the arrangement of the Pd(dmit)$_2$ molecules in EtMe$_{3}$P[Pd(dmit)$_2$]$_2$ with the bars representing the Pd(dmit)$_2$ molecules   viewed along their long axes. (b) Modeling of the in-plane electronic structure by an isosceles triangular lattice with two different transfer integrals $t$ and $t'$ as discussed in ref.\,\onlinecite{Itou2008,Scriven2012}. Each dot represents a [Pd(dmit)$_2$]$_2^{-}$ dimer (dotted circles in (a)). }
\end{figure}

Correlated electrons on geometrically frustrated lattices give rise to complex magnetic behavior. Prominent examples can be found in the series of organic charge-transfer salts Et$_x$Me$_{4-x}Z$[Pd(dmit)$_2$]$_2$ where Et = C$_2$H$_5$, Me = CH$_3$, dmit = 1,3-dithiole-2-thione-4,5-dithiolate, x = 0, 1, 2 and $Z$ = P, As, Sb \cite{Kato2004}. In these compounds strongly dimerized [Pd(dmit)$_2$]$_2^-$ anions form a two-dimensional (2D), slightly distorted triangular lattice (cf.\,Fig.\,\ref{structure}) which alternates with layers of charge-compensating (EtMe$_3Z$)$^{+}$ cations. At ambient pressure and low temperatures the systems are Mott insulators with one unpaired electron residing on each dimer \cite{Tamura2004} resulting in a 2D spin $S$ = 1/2 triangular lattice. Of particular current interest have been the compounds EtMe$_{3}$P[Pd(dmit)$_2$]$_2$ and EtMe$_{3}$Sb[Pd(dmit)$_2$]$_2$. The $Z$ = Sb system lacks long-range magnetic order down to temperatures as low as 20\,mK \cite{Itou2010} and has been considered as a very promising candidate for a quantum-spin liquid (quantum-SL) \cite{Itou2008, Kanoda2011}. On the other hand, for the $Z$ = P system, a phase transition from a high-temperature paramagnetic phase into a spin-gapped valence-bond-solid (VBS) state has been observed below $T_{VBS}$ = 25\,K \cite{Tamura2006}. The VBS state is characterized by the formation and static ordering of spin-singlet valence bonds as a result of lattice deformations and an accompanied alternation in the dimer-dimer interactions.\\ 

It was found that for both compounds the high-temperature paramagnetic regime \cite{Tamura2006,Itou2008}, covering temperatures around and above the broad maximum in the susceptibility at $T_{max}^{\chi} \simeq$ 70\,K ($Z$ = P) and 50\,K (Sb), can be reasonably well described by a 2D $S$ = 1/2 Heisenberg antiferromagnet on a triangular lattice with an exchange coupling constant $J$/$k_B \approx$ 250\,K for $Z$ = P and 220-250\,K for $Z$ = Sb. Electronic structure calculations based on extended H\"{u}ckel molecular orbitals \cite{Itou2008} and density functional theory (DFT) \cite{Scriven2012, Tsumuraya2013} indicate, however, that the interdimer interactions are anisotropic. The DFT calculations in ref.\,\onlinecite{Scriven2012} suggest a description in terms of a Heisenberg model on an anisotropic triangular lattice with two different transfer integrals $t'$ and $t$ (see Fig.\,\ref{structure} for a definition of $t$ and $t'$ ) with a ratio $t'$/$t$ = 0.87 for $Z$ = P and 0.79 for $Z$ = Sb corresponding to a degree of frustration of $J'$/$J$ = ($t'$/$t$)$^{2}$ = 0.75 ($Z$ = P) and 0.62 (Sb). It has been argued in ref.\,\onlinecite{Scriven2012} that with these numbers, both compounds fall into the range 0.5 $\lesssim J'/J \lesssim$ 0.9 where antiferromagnetic order is no longer a very stable ground state of the triangular lattice. Rather it has been suggested that for these $J'$/$J$ ratios perturbative terms in the Hamiltonian such as intradimer dynamics, ring exchange, elastic forces or differences in the crystal structure can be decisive for the ground state of the material. In fact, a structural peculiarity of the VBS system EtMe$_3$P[Pd(dmit)$_2$]$_2$ is the uniform stacking direction of the Pd(dmit)$_2$ molecules \cite{Kato2006}. This contrasts with the usual $\beta'$ structure of the [Pd(dmit)$_2$]-based salts \cite{Kato2004}, also shared by the quantum-SL system EtMe$_3$Sb[Pd(dmit)$_2$]$_2$, where the stacking direction of the Pd(dmit)$_2$ molecules alternates from layer to layer. It has been speculated in ref.\,\onlinecite{Tamura2006} that it is this structural difference which could be of relevance for the different ground states of the $Z$ = P and Sb systems.\\

The VBS transition in EtMe$_3$P[Pd(dmit)$_2$]$_2$ was studied by means of X-ray structural investigations \cite{Tamura2006}. At room temperature the system has space group $P2_1$/$m$ (monoclinic) and structural parameters $a$ = 6.3960(3), $b$ = 36.691(1), $c$ = 7.9290(3)\,${\AA}$, $\beta$ = 114.302(2)\,$^{\circ}$, $V$ = 1695.9(1)\,${\AA}^{3}$ and Z = 2, with the face-to-face stacking of the [Pd(dmit)$_2$]$_2$-units along the $c$-axis, see Fig.\,\ref{structure}. The structure at 10\,K, \emph{i.e.}, in the VBS state, has also $P2_1$/$m$ symmetry (monoclinic) with $a$ = 6.3270(2), $b$ = 36.536(1), $c$ = 14.2620(5)\,${\AA}$, $\beta$ = 90.552(3)\,$^{\circ}$, $V$ = 3296.7\,${\AA}^{3}$ and Z = 4. Note that in the VBS state the $c$-axis is defined as \textbf{$c$} = 2\textbf{$c_{0}$} + \textbf{$a_0$}, where \textbf{$c_0$} and \textbf{$a_0$} are the $c$- and $a$-axes of the high-temperature structure. Due to the twofold alternation of the interdimer distances, the periodicity along the stacking direction has doubled. In addition, it was found that upon cooling from 28\,K down to 10\,K, \emph{i.e.}, through $T_{VBS}$, the lattice expands slightly along the $a$ and $b$ directions, while no expansion was observed along the stacking $c$ direction \cite{Tamura2006}.\\

In this work we report measurements of the uniaxial thermal expansion coefficients on the title compound over a wide range of temperature. Particular attention is payed to the phase transition into the nonmagnetic VBS ground state. Because of the extraordinarily high sensitivity of thermal expansion measurements with regard to both resolution of length changes and variations in temperature, these studies complement existing structural information based on X-ray investigations on some fixed temperatures, and, by this, enable to gain a deeper insight into the subtle structural changes preceding and accompanying this transition.\\

\section{Experiment}

The high-quality single crystals studied in this work were prepared by air oxidation of (EtMe$_{3}$P)$_{2}$[Pd(dmit)$_2$] in acetone containing acetic acid at 5-10$^{\circ}$C for 3-4 weeks. The high quality of the crystal is guaranteed by an X-ray crystal structure analysis with an R-factor of 0.0307 and a GOF (goodness of fit) of 1.127. The monoclinic $P2_1$/$m$ phase was obtained as the main phase, accompanied by triclinic- and $\beta'$-phases (monoclinic C2/c) as minor phases. Experiments have been performed on two different single crystals taken from the same batch (no.\,839) having dimensions of typically 1.3$\times$0.8$\times$0.1\,mm$^3$. The thermal expansion measurements were performed by using an ultrahigh-resolution capacitive dilatometer, built after \cite{Pott1983}, enabling the detection of length changes $\Delta$$l$ $\geq$ 10$^{-2}$ \AA, where $l$ is the length of the sample. For the measurements along the out-of-plane $b$-axis, where the crystal size is only 0.1\,mm, a stack of two crystals, linked together by a tiny amount of Apiezon N grease (corresponding to a $\mu$m layer thickness), was used to increase the signal-to-noise ratio. The measurements were performed upon heating and cooling with a slow sweep rate of $\pm$ 1.5\,K/h ($T$ $<$ 35\,K) to ensure thermal equilibrium. \\

\section{Results and Discussion}

\begin{figure}[!htb]
\centering
\includegraphics[width=1\columnwidth]{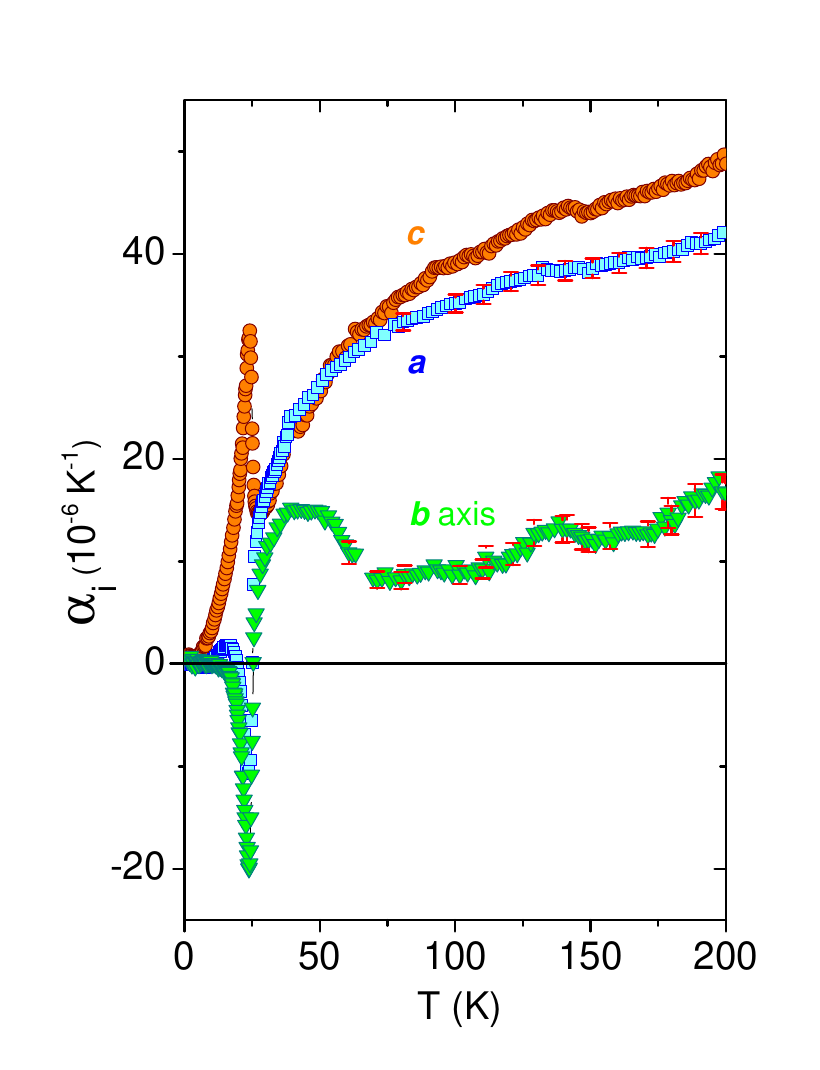}
\caption{\label{all-axes} (color online). Temperature dependence of the uniaxial expansivities $\alpha$$_i$ of EtMe$_3$P[Pd(dmit)$_2$]$_2$ measured along the in-plane \emph{i} = \emph{a}- and \emph{c}-axis and along the out-of-plane \emph{b}-axis. Crystal \#1 and \#2 were used for measurements of $\alpha_c$ and $\alpha_a$, respectively, while a stack of two pieces of crystal \#2 was used for measurements of $\alpha_b$.}
\end{figure}

Figure \ref{all-axes} shows the uniaxial expansion coefficients $\alpha$$_i$ = $l_i^{-1}$d$l_i$/d$T$ (\emph{i = a, b, c}) measured along the in-plane \emph{a}- and \emph{c}-axis and the out-of-plane \emph{b}-axis below 200\,K. The salient features include (i) the distinct and strongly anisotropic phase transition anomalies in $\alpha$$_i$ at $T_{VBS}$, which will be discussed below in more detail, and (ii) a pronounced in-plane versus out-of-plane anisotropy for temperatures $T > T_{VBS}$ together with (iii) a highly anomalous out-of-plane expansivity $\alpha_{b}$. While the in-plane expansion coefficients $\alpha_{c}$ and $\alpha_{a}$ for $T > T_{VBS}$ grow rapidly upon warming, with the tendency to level off at high temperatures, the out-of-plane expansion coefficient $\alpha_{b}$ reveals a maximum around 40\,K followed by a broad minimum around 100\,K above which it slowly increases. These expansivities in the paramagnetic regime $T > T_{VBS}$ are remarkable in two respects. First, the in-plane expansion coefficients, reminiscent of an ordinary, phonon-dominated behavior, characterized by a monotonous increase upon warming and saturation at high temperatures, are only weakly anisotropic. This appears surprising at first glance given the uniform orientation of the Pd(dmit)$_2$ molecules (cf.\,Fig.\,\ref{structure}) from which an anisotropic contribution to the in-plane expansivity can be expected. Second, the distinct deviations from such a monotonously increasing behavior in $\alpha_{b}$, featuring a distinct maximum around 40\,K, indicates substantial non-phononic contributions for $T > T_{VBS}$. \\

The weak anisotropy in $\alpha_{a}$ and $\alpha_{c}$ for $T > T_{VBS}$ suggests that in this temperature range the contribution of the Pd(dmit)$_2$ molecules to the expansivity is rather small. This implies that the in-plane expansivities are dominated by anharmonic motions of the EtMe$_{3}$P counterions. A dominant contribution of these isolated, weakly bound molecular units is consistent with recent results obtained on the metallic (TMTSF)$_2$PF$_6$ salt \cite{Foury2013}. There it was found that the thermal expansion coefficient is dominated by librational motions of the PF$_6$ units \cite{Foury2013} which could be well described by Einstein oscillators. Thus, for the present case, a significant contribution of these Einstein-like local modes of the EtMe$_3$P cations to the lattice expansivity can be expected. For an adequate description of the whole lattice effects, however, the contributions of the Pd(dmit)$_2$-derived molecular modes should be considered as well. The contribution of these dispersive modes are expected to follow a Debye function.\\

\begin{figure}[!htb]
\centering
\includegraphics[width=\columnwidth]{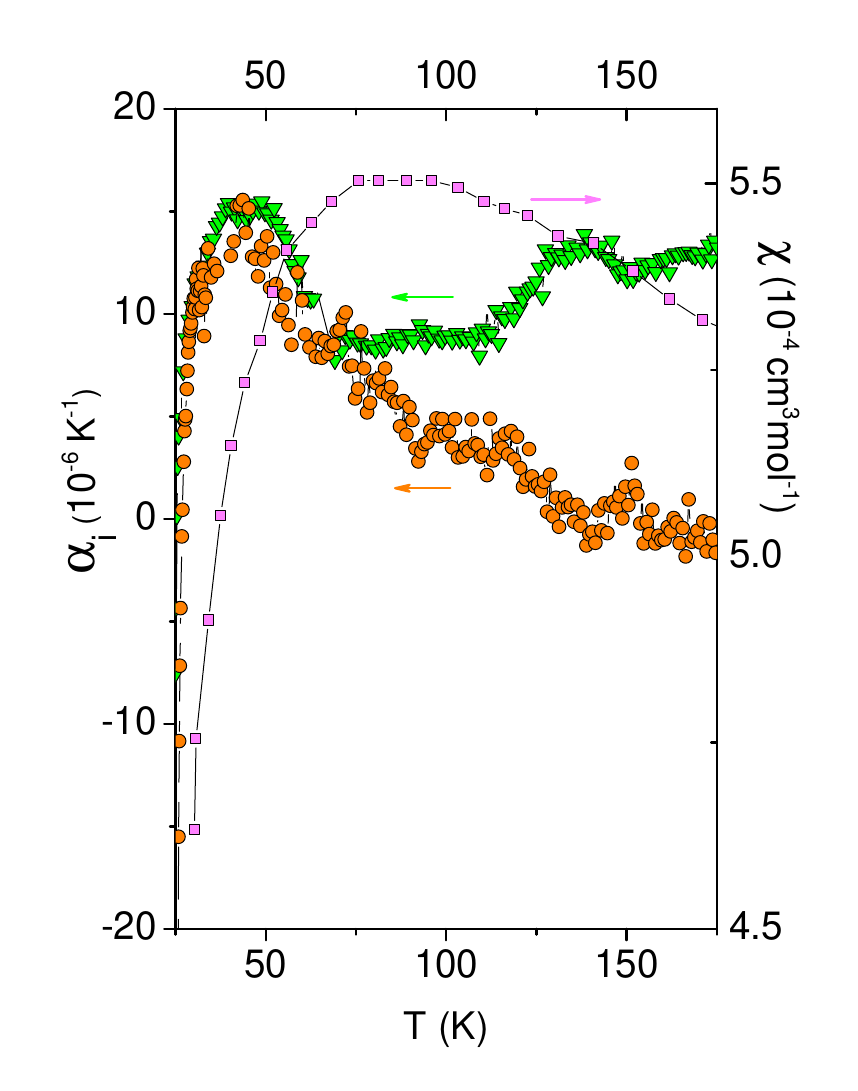}
\caption{\label{Debye} (color online). Combination of in-plane expansivities $\alpha_c$ - 1.15 $\cdot \alpha_a$ multiplied by -2.8 (orange circles, left scale) to get rid of the unknown phonon contribution, see text. For comparison, the out-of-plane data $\alpha_b$ data (green inverted triangles, left scale) and a blowup of magnetic susceptibility data \cite{Tamura2006} (magenta squares, right scale) are shown for the same temperature range.}
\end{figure}

However, attempts to model the lattice expansivity by a combination of an Einstein and a Debye contribution, with characteristic temperatures, $\Theta_E$ and $\Theta_D$, respectively, and two independent Gr\"{u}neisen parameters $\gamma_E$ and $\gamma_D$ as prefactors, were unsuccessful due to the limited temperature range for the fit and the large number (four) of free parameters. Instead, we propose a different way for analyzing the in-plane data. We assume that the in-plane expansivities each consist of a lattice and a magnetic contribution: $\alpha_a$ = $\alpha^{lat}_a$ + $\alpha^{mag}_a$ and $\alpha_c$ = $\alpha^{lat}_c$ + $\alpha^{mag}_c$. We further make the reasonable assumption that scaling relations $\alpha^{lat}_c$ = $A\cdot \alpha^{lat}_a$ and $\alpha^{mag}_c$ = $B\cdot \alpha^{mag}_a$ hold, where the proportionality constants $A$ and $B$ account for differences in the uniaxial Gr\"{u}neisen parameters for the $a$ and $c$-axis, \emph{i.e.}, differences in the uniaxial compressibilities and uniaxial pressure dependencies of the characteristic energy scales. Moreover, we assume that the magnetic contributions at higher temperatures $T \gtrsim$ 135\,K are small so that the expansivities here can be considered to reflect the pure lattice effects. In fact, by multiplying the $\alpha_a$ data by a factor 1.15 (= $A$), we find a collapse of both data sets for $T \gtrsim$ 135\,K within the experimental uncertainty. Therefore, we expect the quantity $\alpha_c$ - 1.15$\cdot \alpha_a$ = $\alpha^{mag}_c$ - $1.15\cdot \alpha^{mag}_a$, to be, to a good approximation, free from the unknown lattice contribution, and to reflect the magnetic contribution except for an unknown prefactor. In Fig.\,3 we show the so-derived quantity $\alpha_c$ - 1.15$\cdot \alpha_a$ multiplied by a factor -2.8. As the figure clearly demonstrates, the magnetic contribution to the in-plane expansivities, $\alpha^{mag}_a$ and $\alpha^{mag}_c$, reveal an anomaly around 40\,K, the temperature dependence of which matches the one observed in the out-of-plane data $\alpha_b$ in great detail.\\

The occurrence of more or less pronounced maxima (or minima) in the uniaxial expansivities at $T_{max}^{\alpha}$, coinciding with a maximum in the specific heat at $T_{max}^{C}$, are known from low-dimensional quantum-spin systems, see, \emph{e.g.},\,refs.\,\onlinecite{Winkelmann1995, Bruhl2007, Manna2010}. These extrema in $\alpha_{i}(T_{max}^{\alpha}$) reflect the strain dependence of the relevant magnetic coupling energy and are thus often found to be strongly anisotropic. Those anomalies are accompanied by a maximum in the magnetic susceptibility at a temperature $T_{max}^{\chi}$ which may differ from $T_{max}^{\alpha}$ = $T_{max}^{C}$. For example, a ratio $T_{max}^{\chi}$/$T_{max}^{C}$ = $T_{max}^{\chi}$/$T_{max}^{\alpha}$ = 1.34 is obtained for the uniform antiferromagnetic spin-1/2 Heisenberg chain, see, \emph{e.g.}, ref.\,\onlinecite{Hammar1999, Rohrkamp2010} for specific heat and thermal expansion measurements, and 3.08 for the alternating exchange variant with an alternation parameter $\delta$ = 0.16 \cite{Buhler2001}. For even stronger alternation and by introducing frustration due to next-nearest-neighbor interactions, $T_{max}^{\chi}$/$T_{max}^{C}$ further increases slightly to 3.6. On the other hand for 2D strongly frustrated triangular-lattice spin-1/2 Heisenberg antiferromagnets, such as Cs$_{2}$CuBr$_{4}$ with a ratio $J'$/$J$ = 0.74 \cite{Ono2004} or the spin liquid candidate $\kappa$-(BEDT-TTF)$_{2}$Cu$_{2}$(CN)$_{3}$ with $J'$/$J$ = 0.64 - 0.74 \cite{Jeschke2012}, one obtains a ratio $T_{max}^{\chi}$/$T_{max}^{C}$ close to one \cite{Ono2005, Manna2010, Shimizu2007}.\\

For the present material, where the susceptibility reveals a broad maximum around 70\,K (= $T_{max}^{\chi}$) due to short-range magnetic correlations, we do in fact expect to observe a corresponding feature in $\alpha_{i}$. What is remarkable, however, is the fact that the anomalies revealed in $\alpha_{i}$ are located at a temperature $T_{max}^{\alpha}$ distinctly lower than $T_{max}^{\chi}$, with $T_{max}^{\chi}$/$T_{max}^{\alpha} \simeq$ 70\,K/40\,K = 1.75 $\pm$ 0.1, \emph{i.e.}, distinctly above one. This might be an indication that the system is in fact better described by three distinct coupling constants $J$, $J'$ and $J''$, corresponding to three distinct transfer integrals, $t$, $t'$, $t''$ in Fig.\,\ref{structure}, with one dominant coupling, so that a quasi-one-dimensional scenario would be more appropriate.\\

\begin{figure}[!htb]
\centering
\includegraphics[width=\columnwidth]{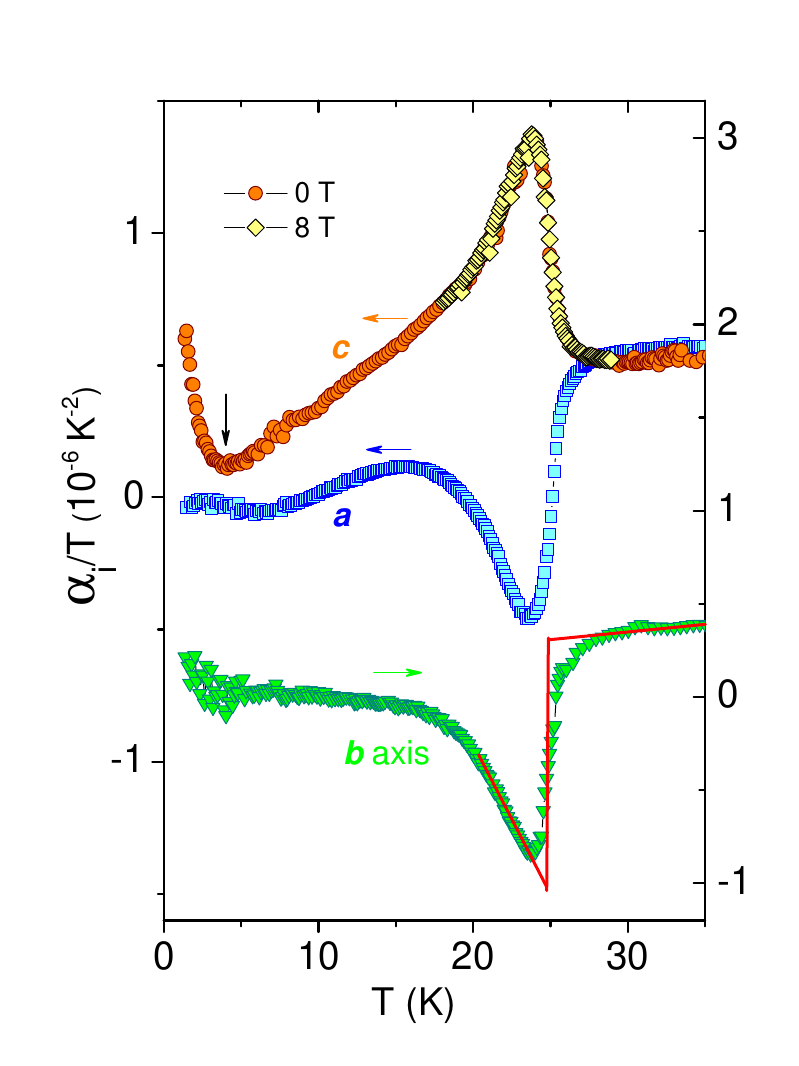}
\caption{\label{Fig06} (color online). Plot of the out-of-plane uniaxial expansivity $\alpha_{b}$ as $\alpha_{b}$/$T$ \emph{vs} $T$. The equal-areas construction for replacing the broadened phase transition anomaly by an idealized sharp one is exemplarily shown for $\alpha_{b}$. Note the different scales used for the in-plane and out-of-plane data. The arrow around 3.5\,K marks the position of the minimum in $\alpha_c$/$T$.}
\end{figure}

In order to analyze the lattice effects at and below $T_{VBS}$ in more detail, we show in Fig.\,\ref{Fig06} the quantities $\alpha_{i}$ as $\alpha_{i}$/$T$ on expanded scales. For all three uniaxial expansivities we find a well-pronounced, slightly broadened $\lambda$-type phase transition anomaly around 25\,K which is assigned to the second-order phase transition into the low-$T$ VBS phase reported in the literature \cite{Tamura2006, Shimizu2007, Itou2009}. Measurements upon warming and cooling (not shown) with a slow rate of $\pm$1.5\,K/h across the phase transition failed to detect any hysteresis, consistent with the second-order character of the phase transition. Measurements of $\alpha_i$ for $i$ = $a$, $b$ and $c$ in magnetic fields up to $B$ = 8\,T were found to have no effect on the phase transition, see, \emph{e.g.}, the data for $\alpha_c$ taken at $B$ = 8\,T in Fig.\,\ref{Fig06}. In order to determine the discontinuities at $T_{VBS}$, defined as $\Delta \alpha$ = lim$_{T\rightarrow T_{VBS}}$ [$\alpha(T<T_{VBS}) - \alpha (T>T_{VBS})$], we replace the slightly broadened anomalies by idealized sharp ones in an equal-areas construction in an $\alpha$/$T$ \emph{vs} \,$T$ plot, exemplarily shown in Fig.\,\ref{Fig06} for $\alpha_{b}$. This yields $\Delta \alpha_{a}$ = -(27$\pm$1.5)$\cdot$10$^{-6}$K$^{-1}$, $\Delta\alpha_{c}$ = +(25$\pm$1.5)$\cdot$10$^{-6}$K$^{-1}$ and $\Delta\alpha_{b}$ = -(32$\pm$2)$\cdot$10$^{-6}$K$^{-1}$ and a transition temperature $T_{VBS}$ = (24.9$\pm$0.2)\,K. Note that $T_{VBS}$ coincides with the transition temperature derived from magnetic measurements \cite{Tamura2004}.\\

Focussing first on the in-plane effects at $T_{VBS}$. The positive discontinuity in $\alpha_{c}$ means that upon cooling through $T_{VBS}$ the $c$-axis -- the stacking axis of the dimers, cf.\,Fig.\,\ref{structure} -- shrinks. This is accompanied by an expansion of the $a$-axis by the same relative amount. As a result the formation of the spin-singlet valence bonds is accompanied by a significant deformation of the in-plane triangular lattice and, by this, changes in the hopping terms (exchange couplings) along the $a$ and $c$-axes. Given that the hopping probability between adjacent molecules increases by decreasing their distance one may expect that cooling into the VBS state is accompanied by an increase in the transfer integral along the $c$-axis and a decrease along the $a$-axis.\\

Remarkably, the largest anomaly is observed along the cross-plane \emph{b}-axis which also expands (negative $\Delta \alpha_b$) upon cooling. This appears rather unexpected at first glance given the pronounced anisotropy of the phonon contributions, with $\alpha_{b}^{lat}$ $\ll$ $\alpha_{a}^{lat}$, $\alpha_{c}^{lat}$, indicating a rather small $b$-axis compressibility. We stress, however, that the lattice effects for $T > T_{VBS}$ are dominated by the librational motions of the EtMe$_{3}$P counterions which are likely to be less involved in the phase transition. On the other hand, the lattice contribution of the Pd(dmit)$_2$ molecules, to which the spins are expected to couple most strongly, could not be determined quantitatively. The observation of a pronounced out-of-plane effect at $T_{VSB}$ may thus indicate that their contribution to the lattice expansivity is in fact strongest along the $b$-axis. Since the formation of a VBS state is believed to be closely related to the in-plane magnetic correlations of the material, one may speculate that the expansion of the $b$-axis lattice parameter results from a shear deformation in the position of neighboring Pd(dmit)$_2$ molecules along the out-of-plane direction or a tilt of these molecules in response to the in-plane lattice deformations.\\

According to the Ehrenfest relation, the discontinuities $\Delta \alpha_i$ can be used to calculate the uniaxial pressure dependencies of $T_{VBS}$ in the limit of vanishingly small pressure $p_i$ applied along the $i$ axis:

\begin{equation}\label{Ehrenfest}
 ( \frac{\partial T_{VBS}}{\partial p_i})_{p_i \rightarrow 0} = T_{VBS}\cdot V_{mol}\cdot \frac{\Delta \alpha_i}{\Delta C}.
\
\end{equation}

Here $\Delta C$ is the discontinuity in the specific heat at $T_{VBS}$ and $V_{mol}$ = 4.964$\cdot$10$^{-4}$m$^{3}$mol$^{-1}$ is the molar volume. By using $\Delta C$ = 9.8 J mol$^{-1}$K$^{-1}$\cite{Kato2013} we find ($\partial T_{VBS}/\partial p_{a})_{p_a \rightarrow 0}$ = -(0.34$\pm$0.03)\,K/GPa, ($\partial T_{VBS}/\partial p_{c})_{p_c \rightarrow 0}$ = +(0.32$\pm$0.03)\,K/GPa and ($\partial T_{VBS}/\partial p_{b})_{p_b \rightarrow 0}$ = -(0.42$\pm$0.05)\,K/GPa. From the sum of these uniaxial pressure coefficients we can calculate the pressure dependence under hydrostatic pressure of ($\partial T_{VBS}/\partial p)_{p \rightarrow 0}$ = -(0.44$\pm$0.1)\,K/GPa. This value is somewhat larger than the pressure coefficient of -0.27\,K/GPa as read off the phase diagram in ref.\,\onlinecite{Shimizu2007}. Besides providing an independent measurement of the pressure dependence of the VBS state, the present study discloses a rather unexpected result: the strong suppression of $T_{VBS}$ under hydrostatic pressure is solely due to the out-of-plane uniaxial pressure component as the in-plane pressure effects just cancel each other out.\\

Apart from the phase transition at $T_{VBS}$, the thermal expansion data in Fig.\,\ref{Fig06} reveal indications for another anomaly below $T_{VBS}$. Our data uncover a distinct minimum in $\alpha_c$/$T$ around 3.5\,K. We note that around the same temperature a broad peak was observed in the spin-lattice relaxation rate $T_1^{-1}$ derived from NMR measurements \cite{Itou2009}. This effect has been attributed to an inhomogeneous relaxation in the system due to a minority of remaining unpaired free spins. \\

In summary, measurements of the uniaxial expansivities on EtMe$_{3}$P[Pd(dmit)$_2$]$_2$ reveal anomalous behavior above and around the phase transition at $T_{VBS}$ = 25\,K into the low-temperature valence-bond-solid phase. The pronounced in-plane \emph{vs} out-of-plane expansivity in the paramagnetic regime is attributed to the dominant contribution of librational oscillations of the EtMe$_{3}$P counterions. At a temperature $T_{max}^{\alpha} \approx$ 40\,K a distinct anomaly in the uniaxial expansivities is found which is assigned to short-range antiferromagnetic correlations giving rise also to a maximum in the magnetic susceptibility at $T_{max}^{\chi}$ around 70\,K as reported in the literature. From the ratio $T_{max}^{\chi}$/$T_{max}^{\alpha}$ = 1.75, being distinctly larger than 1 as observed for quasi-2D slightly anisotropic triangular-lattice spin-1/2 antiferromagnets, it is conjectured that the present material may be better described by a quasi-one-dimensional scenario, \emph{i.e.}, a triangular lattice with three distinctly different hopping terms $t$, $t'$ and $t''$. The discrepancy to the DFT calculations \cite{Scriven2012}, favoring a description in terms of 2D triangular lattice with two transfer integrals $t$ and $t'$, may thus indicate uncertainties in the structural data on which these calculations are based.\\

Measurements of $\alpha_i$ around the phase transition into the VBS state reveal well-pronounced, slightly broadened $\lambda$-like phase transition anomalies. The data show that cooling into the VBS state is accompanied by a contraction of the in-plane $c$-axis lattice parameter, along which the valence bonds form, and an elongation of the in-plane $a$-axis by the same relative amount. Surprisingly, the strongest response is found along the out-of-plane $b$-axis which also expands upon cooling through $T_{VBS}$. As a consequence, the strong suppression of $T_{VBS}$ found under hydrostatic pressure is solely due to the out-of-plane pressure component while the in-plane pressure component has no effect on $T_{VBS}$.\\

\section{Acknowledgments}
We acknowledge financial support by the Deutsche Forschungsgemeinschaft via the SFB/TR49 and the Grant-in-Aid for Scientific Research (S)(No. 22224006) for the Japan Society for the Promotion of Science (JSPS).

\end{document}